\documentclass[12pt]{article}

\usepackage{comment}
\usepackage{tikz}
\usepackage{caption}

\usepackage{bm}

\usepackage{mathtools}

\usepackage[mathscr]{euscript}

\usepackage{slashed}

\usepackage{amsmath}

\date{}

\title{\textbf{Symplectic Quantization and General Constraint Structure of a Prototypical Second-Class System
}}
\author{ \textbf{Ignacio S. Gomez$^{a,b}$, Vipul Kumar Pandey$^{c}$ and
Ronaldo Thibes$^{a,b}$}
\\\\
\textit{$^{a}$\small{\!Universidade Estadual do Sudoeste da Bahia}}\\
\textit{\small{DCEN, Rodovia BR 415, Km 03, S/N, Itapetinga -- 45700-000, Brazil}}
\\\\
\textit{$^{b}$\small{\!PROFÍSICA -- Programa de Pós-Gradua\c{c}\~ao em F\'\i sica}}\\
\textit{\small{Universidade Estadual de Santa Cruz, 45650-000 Ilhéus, Bahia, Brazil}}
\\\\
\textit{$^{c}$\small{\!Department of Physics, Chandigarh University}}\\
\textit{\small{Mohali -- 140413, India}}
}

\usepackage{graphicx}
\usepackage{amssymb,amsfonts,amssymb,amsthm}
\usepackage{amsmath}

\begin{document}

\maketitle

\abstract{
We discuss a general prototypical constrained Hamiltonian system with a broad application in quantum field theory and similar contexts where dynamics is  defined through a functional action obeying a stationarity principle.  The prototypical model amounts to a Dirac-Bergmann singular system, whose constraints restrict the actual dynamics to occur within a differential submanifold, as is the case in the major part of field theoretical models with gauge symmetry.  We apply the Dirac-Bergmann algorithm in its full generality unraveling a total of $4m$ second-class constraints and obtain the corresponding Dirac brackets algebra in phase space.  We follow with the Faddeev-Jackiw-Barcelos-Wotzasek approach in which the geometric character of the mentioned submanifold is emphasized by means of an internal metric function encoding its symplectic properties.  We consider two straightforward examples, applying our general results to constrained motion along a toroidal geometry and to a Lorentz violating toy model in field theory.  Since toroidal geometry has been recently used in cosmological models, we suggest how our results could lead to different proposals for the shape of the universe in cosmology.
}

\section{Introduction}
The theory of constrained Hamiltonian systems has been systematically employed as an important tool for the quantization process of central models in physics appearing in classical and quantum mechanics, field theories, gravity, string and brane theories and related frameworks defined in terms of a minimal action principle \cite{Sundermeyer:1982gv, Gitman:1990qh, Henneaux:1992ig}.  Whether one reasons by canonical or functional quantization methods, the existence of Dirac-Bergmann (DB) constraints permeates the analysis of the most fundamental mechanisms present in modern physics.  As a clear attesting fact, we mention that all gauge invariant theories represent instances of DB constrained systems \cite{Thibes:2020jfp}.  Due to the huge amount of different constructs and approaches to quantum field theory, it is not rare to see the same constraint structure repeating itself in several disguised forms throughout seemingly different physical systems and contexts.  
It would be surely helpful to understand those structures in a model independent way.
In the present work, we discuss a prototypical second-class DB system with a fairly general constraint structure which can be seen in an extensive amount of both mechanical and field theory contexts.  The detailed analysis of its various classical and quantum aspects leads to a comprehensive knowledge which can be applied to particular models enhancing similarities, producing general insights and practical shortcuts for otherwise long unnecessary calculations.  In particular, in the present work, we explicitly obtain the general final expressions for the correct bracket structure in phase space, both {\it a la} Dirac as well as in terms of the more modern symplectic Faddeev-Jackiw-Barcelos-Wotzasek (FJBW) approach.  In fact, the FJBW approach has the advantage of producing the proper quantum commutation relations directly from the Faddeev-Jackiw brackets and highlights the system internal geometric structure smoothly realized within a minimal submanifold characterized by the relevant physical variables. 

The prototypical system (PS) discussed here comes from a generalization of particular cases previously studied in references 
\cite{Thibes:2020yux, Pandey:2021myh, Mandal:2023pdk, Neto:2022jgl}.  It actually constitutes a Dirac-Bergmann singular system in which the constraints are described by holonomic functions fully exploited in terms of Lagrange multipliers considered as natural internal variables.  A first specific realization was introduced in  \cite{Thibes:2020yux}, where the corresponding BRST symmetries were fully obtained followed by its functional BFV quantization.  In \cite{Pandey:2021myh}, after some refinement, the  BFFT formalism was applied in a fairly general way converting the constraints to first-class prior to quantization, shedding light to important issues on gauge symmetry generation in quantum field theory, where we can usually see such conversions in an {\it ad hoc} manner.   It was also shown in \cite{Pandey:2021myh} that the proposed PS is able to describe Lorentz violating scenarios associated to possible consistent extensions of the standard model
\cite{Colladay:1998fq, Kostelecky:2003fs}.  In \cite{Mandal:2023pdk}, various forms of BRST and BRST-related symmetries were explored and worked out throughout the PS, where it was shown that it constitutes a very useful framework for studying and interrelating those quantum symmetries in a unified form with connections to common gauge-fixings employed in QED and QCD. 
In \cite{Neto:2022jgl}, important hidden symmetries within the PS were revealed under aspects coming from the modified gauge-unfixing formalism \cite{Neto:2006gt, Costa:2023xwj}, where a corresponding gauge-invariant Hamiltonian was obtained, with the main results exemplified in the nonlinear sigma model.

In the present work, we considerably extend that initial system and deeper explore its internal algebraic properties aiming to enhance its applicability not just to traditional field theory but to other modeling contexts.  As a typical example, we mention the recent work of  Nejad, Dehghani and Monemzadeh \cite{Nejad:2015rfa} in cosmology, in which the authors discuss toroidal geometry as an interesting possibility for modeling a universe with extra dimensions coming from Lagrange multipliers and Wess-Zumino fields.  We shall show here that not only is it perfectly feasible to reproduce the symplectic structure discussed in \cite{Nejad:2015rfa} from our PS, with considerable technical advantages, but that its higher generality certainly provides room for deviations from a strictly toroidal geometry opening roads for other tentative shapes and geometries for the universe in that context.  With the consistent general analysis and formulae developed here, one is ready to tackle those and similar problems from a broader viewpoint seeing the forests for the trees.

With that in mind, for the reader's convenience, we have organized our line of arguments as follows.  In Section {\bf 2} below, we introduce the PS as a dynamical system defined within a $2n$-dimensional phase space subjected to $m$ holonomic conditions reducing its effective dynamics to a lower-dimensional submanifold.  In Section {\bf 3}, we apply the DB algorithm, extracting a total of $4m$ constraints to be fully classified under Dirac's nomenclature according to its properties under Poisson brackets.  We find the second-class condition in terms of the determinant of a key matrix constructed from the derivatives of the constraints and compute the complete Dirac brackets algebra in the extended phase space.  In Section {\bf 4}, turning to the geometric FJBW approach, we perform the general PS symplectic analysis, in which we see that the inversibility condition for the main symplectic matrix coincides with the previous second-class condition, and compute the Faddeev-Jackiw brackets in its most general form.  In Section {\bf 5}, we discuss two specific applications: the first one related to the above mentioned toroidal universe model and the second one related to a Lorentz violating bumblebee field theory.  We end in Section {\bf 6} with our conclusion and final comments.

\section{The Prototypical Second-Class System}
Let $T_\alpha(q^i)$, $\alpha=1,\dots,m$, denote a set of $m$ holonomic conditions to be imposed along the evolution of a dynamical system under the action of a physical potential $V(q^i)$, depending on $n$ generalized coordinates $q^i$, $i=1,\dots,n$, with $n>m$.  The actual realization is achieved by introducing $m$ additional independent coordinates $l^\alpha$ and constructing the Lagrangian function\footnote{Repeated index summation, with $i,j,k=1,\dots,n$ and $\alpha,\beta,\gamma=1,\dots,m$, is always implied.}
\begin{equation}\label{L}
L(l^\alpha,q^k,{\dot{q}}^k)=\frac{1}{2}f_{ij}(q^k){{\dot{q}}^i}{{\dot{q}}^j}
-V(q^k)
-l^\alpha T_\alpha(q^k)
\,,
\end{equation}
where $f_{ij}(q^k)$ represents a given $n\times n$ symmetric tensor with a corresponding inverse $f^{ij}(q^k)$ satisfying 
\begin{equation}\label{2}
f^{\,ik}\,f_{kj} = f_{jk}\,f^{\,ki} = \delta^{\,i}_{\,j}\,.
\end{equation}
This simple system captures the canonical constraints structure of many important field theory models spread throughout the literature. 
The main reason for that stems from the fact that
the Lagrangian function (\ref{L}) represents a Dirac-Bergmann constrained dynamical system, by which we mean it has a degenerated Hessian matrix and the  Legendre transformation leading to its corresponding Hamiltonian description in phase space is not invertible, as is the case in nearly all field theories of interest as well as in many quantum mechanical models viewed as $(0+1)$ field theories.  From the derivatives of the functions $T_\alpha$ with respect to $q^i$, denoted by
\begin{equation}
T_{\alpha,i}\equiv \frac{\partial T_\alpha}{\partial {{q}}^i}
\,,
\end{equation}
we construct further the key symmetric matrix
\begin{equation}\label{w}
w_{\alpha\beta}(q^i)\equiv f^{ij}T_{\alpha, i} T_{\beta,j}
=w_{\beta\alpha}(q^i)
\,,
\end{equation}
which we assume to have a non-null determinant, i.e.,
\begin{equation}\label{assumption}
w\equiv \det w_{\alpha\beta} \neq 0
\,.
\end{equation}
The phase space associated to (\ref{L}) consists of the $2(n+m)$ canonical coordinates $(q^i,p_i)$ and $(l^\alpha,\pi_\alpha)$ where the canonical momenta are defined by
\begin{equation}
p_i\equiv \frac{\partial L}{\partial {\dot{q}}^i}\,\, 
\mbox{ and }\,\,
\pi_\alpha \equiv \frac{\partial L}{\partial {\dot{l}}^\alpha}
\,.
\end{equation}
Since the time derivatives of the $l^\alpha$ variables do not show up in  (\ref{L}), we are led to a initial set of $m$ primary constraints
\begin{equation}\label{Xi0}
\chi_{(0)\alpha} \equiv \pi_\alpha\,,\,\,\,\,\alpha=1,\dots,m
\,.
\end{equation}
The canonical Hamiltonian, defined in the $(2n+m)$-dimensional primary constraints hypersurface, can be written as
\begin{equation}\label{Hc}
H_c=\frac{1}{2}f^{ij}(q^k)p_ip_j+V(q^k)+l^\alpha T_\alpha(q^k)
\,.
\end{equation}

As a necessary preparative step for the system quantization, we need to unravel and classify the complete canonical constraints set and compute the corresponding Poisson and Dirac Brackets algebras.  That can be done by means of the Dirac-Bergmann algorithm, which we apply in the following section.

\section{Dirac-Bergmann Algorithm}
Due to the presence of constraints, the canonical quantization of (\ref{Hc}) cannot be achieved simply by the naive prescription of introducing operators satisfying commutator relations directly obtained from corresponding classical Poisson Brackets.  Rather, we need Dirac Brackets.
In this section, we show that $H_c$ characterizes a second class dynamical system and calculate its Dirac Brackets algebra in phase space.
In terms of the Dirac-Bergmann algorithm \cite{Dirac:1950pj, Anderson:1951ta, Dirac}, we start by imposing the time conservation of the primary constraints (\ref{Xi0}) along the dynamical evolution governed by the Hamiltonian (\ref{Hc}).  Since
\begin{equation}
\left\{ \chi_{(0)\alpha} , H_c \right\}
= - T_\alpha
\,,
\end{equation}
we obtain a second family of $m$ constraints
\begin{equation}\label{Xi1}
\chi_{(1)\alpha}\equiv T_\alpha
\,,
\end{equation}
corresponding to the initial holonomic conditions.  Further time conservation leads to more two constraint families
\begin{equation}\label{Xi2}
\chi_{(2)\alpha} \equiv f^{ij}p_iT_{\alpha, j}
\,,
\end{equation}
and
\begin{equation}\label{Xi3}
\chi_{(3)\alpha} \equiv \frac{1}{2}Q_\alpha^{ij}p_ip_j
-v_\alpha-l^\beta w_{\alpha\beta}
\,,
\end{equation}
with $w_{\alpha\beta}$ defined by (\ref{w}) and the quantities $Q^{ij}_\alpha=Q^{ij}_\alpha(q^k)$ and $v_\alpha=v_\alpha(q^k)$ defined as
\begin{equation}\label{Qij}
Q^{ij}_\alpha(q^k)\equiv
{\left(f^{il}T_{\alpha, l}\right)}_{,k} f^{kj}
+{\left(f^{jl}T_{\alpha, l}\right)}_{,k} f^{ki}
-f^{kl}T_{\alpha, k} f^{ij}_{\,\,\,\,,l}
=Q^{ji}_\alpha
\,,
\end{equation}
\begin{equation}\label{v}
v_\alpha(q^k)\equiv f^{ij}T_{\alpha, i}V_j
\,.
\end{equation}
No more constraints are produced by the Dirac-Bergmann algorithm and we are left with a total of $4m$ constraint conditions in phase space, given by equations  (\ref{Xi0}), (\ref{Xi1}), (\ref{Xi2}) and (\ref{Xi3}), with Greek indexes always running through $\alpha,\beta=1,\dots,m$.  In order to classify the obtained constraints into first or second classes, we compute the Poisson Brackets among all of them and construct a corresponding $4m\times4m$ square matrix as  
\begin{equation}\label{CM}
\Gamma_{(rs)\alpha\beta}\equiv\lbrace\chi_{(r)\alpha}, \chi_{(s)\beta} \rbrace =
\displaystyle
\left[
\begin{array}{cccc}
0 & 0 & 0 & w_{\alpha\beta} \\ 
0 & 0 &w_{\alpha\beta} &E_{\alpha\beta} \\
0 & -w_{\alpha\beta}  & M_{\alpha\beta} & R_{\alpha\beta} \\
-w_{\alpha\beta}   & -E_{\beta\alpha} & -R_{\beta\alpha} & N_{\alpha\beta}
\end{array}
\right]
\,,
\end{equation}
for $r,s=0,\dots,3$, with the short-hand conventions
\begin{equation}
M_{\alpha\beta}\equiv
\lbrace
\chi_{(2)\alpha}, \chi_{(2)\beta}
\rbrace
=-M_{\beta\alpha}
\,,
\end{equation}
\begin{equation}
E_{\alpha\beta}\equiv
\lbrace
\chi_{(1)\alpha}, \chi_{(3)\beta}
\rbrace
\,,\,\,\,\,
R_{\alpha\beta}\equiv
\lbrace
\chi_{(2)\alpha}, \chi_{(3)\beta}
\rbrace
\,,
\end{equation}
and
\begin{equation}
N_{\alpha\beta}\equiv
\lbrace
\chi_{(3)\alpha}, \chi_{(3)\beta}
\rbrace
=-N_{\beta\alpha}
\,.
\end{equation}
The determinant of the constraints matrix (\ref{CM}) depends only on its secondary diagonal and is given by
\begin{equation}\label{w4}
\det \Gamma_{(rs)\alpha\beta} =
w^4
\end{equation}
with $w$ previously defined in equation (\ref{assumption}) as, in its turn, the determinant of $w_{\alpha\beta}$.  Since we are assuming $w\neq0$, the above result (\ref{w4}) shows that the Lagrangian function (\ref{L}) indeed characterizes a prototypical second-class system.  All $4m$ constraints in phase space are second-class.

Next, in order to compute the Dirac Brackets, we need to understand better the structure of the constraint matrix (\ref{CM}) and its inverse.
Besides $w_{\alpha\beta}$, the remaining entries below the secondary diagonal in (\ref{CM}) can be explicitly calculated as
\begin{equation}\label{M}
M_{\alpha\beta}=f^{ij}p_k
\left[
{\left(f^{kl}T_{\alpha, l}\right)}_{,i}T_{\beta, j}
-
{\left(f^{kl}T_{\beta, l}\right)}_{,i}T_{\alpha, j}
\right]
\equiv
p_k M^k_{\alpha\beta} = - M_{\beta\alpha}
\,,
\end{equation}

\begin{equation}\label{DR}
E_{\alpha\beta}=T_{\alpha, i}Q^{ij}_\beta p_j
\,,
~~~
R_{\alpha\beta}
=
p_i p_j R^{ij}_{\alpha\beta}+ f^{ij}T_{\alpha, i} v_{\beta, j} 
+ l^\gamma f^{ij}T_{\alpha, i} w_{\beta\gamma, j}
\,,
\end{equation}
and
\begin{equation}\label{N}
N_{\alpha\beta}= p_i
\left[
\chi_{(3)\alpha,j}Q^{ij}_\beta
-
\chi_{(3)\beta,j}Q^{ij}_\alpha
\right]
=
p_i p_j p_k Q^{ijk}_{\alpha\beta}
+p_i V^i_{\alpha\beta}+p_il^\gamma N^i_{\alpha\beta\gamma}
=-N_{\beta\alpha}
\,,
\end{equation}
with the handy definitions
\begin{equation}\label{D1}
M^k_{\alpha\beta}(q^k)\equiv f^{ij}
\left[
\left(f^{kl}T_{\alpha, l}\right)_{,i}T_{\beta, j}
-
\left(f^{kl}T_{\beta, l}\right)_{,i}T_{\alpha, j}
\right]
=-M^k_{\beta\alpha}
\,,
\end{equation}
\begin{eqnarray}
Q^{ijk}_{\alpha\beta}(q^k)&\equiv&\frac{1}{6}
\left(
Q^{ij}_{\alpha,l}Q^{kl}_\beta + Q^{ik}_{\alpha,l}Q^{jl}_\beta
+Q^{jk}_{\alpha,l}Q^{il}_\beta
-
Q^{ij}_{\beta,l}Q^{kl}_\alpha
\right.\nonumber\\&&\left.
-  Q^{ik}_{\beta,l}Q^{jl}_\alpha
- Q^{jk}_{\beta,l}Q^{il}_\alpha
\right)
= -Q^{ijk}_{\beta\alpha} = Q^{jik}_{\alpha\beta} 
= Q^{ikj}_{\alpha\beta}
\,,
\end{eqnarray}
\begin{equation}
V^i_{\alpha\beta}(q^k)\equiv
v_{\alpha, j} Q^{ij}_\beta
-
v_{\beta, j} Q^{ij}_\alpha
= - V^i_{\beta\alpha}
\,,
\end{equation}
\begin{equation}
N^i_{\alpha\beta\gamma}(q^k)\equiv
w_{\alpha\gamma, j} Q^{ij}_\beta
-
w_{\beta\gamma, j} Q^{ij}_\alpha
=
-N^i_{\beta\alpha\gamma}
\,,
\end{equation}
and
\begin{equation}\label{D5}
R^{ij}_{\alpha\beta}(q^k)\equiv
{1\over2}{\left(f^{ik}T_{\alpha, k}\right)}_{,l}Q^{lj}_\beta
+
{1\over2}{\left(f^{jk}T_{\alpha, k}\right)}_{,l}Q^{li}_\beta
-\frac{1}{2}f^{kl}T_{\alpha, k} Q^{ij}_{\beta, l}
=R^{ji}_{\alpha\beta}
\,.
\end{equation}
In general, $E_{\alpha\beta}$ and $R_{\alpha\beta}$ in (\ref{DR}) do not have a well-defined symmetry as we may rather obtain
\begin{equation}
E_{(\alpha\beta)}= \frac{1}{2}\left( T_{\alpha, i}Q^{ij}_\beta p_j + T_{\beta, i}Q^{ij}_\alpha p_j
\right)
\,,
\end{equation}
\begin{equation}
E_{[\alpha\beta]}=  \frac{1}{2}\left( T_{\alpha, i}Q^{ij}_\beta p_j - T_{\beta, i}Q^{ij}_\alpha p_j
\right)
\,,
\end{equation}
\begin{equation}
R_{(\alpha\beta)}= p_i p_j R^{ij}_{(\alpha\beta)} + \frac{1}{2}
f^{ij}\Big[
T_{\alpha,i}\left(v_{\beta,j} + l^\gamma w_{\beta\gamma,j}\right)
+
T_{\beta,i}\left(v_{\alpha,j} + l^\gamma w_{\alpha\gamma,j}\right)
\Big]
\,,
\end{equation}
and
\begin{equation}
R_{[\alpha\beta]}= p_i p_j R^{ij}_{[\alpha\beta]} + \frac{1}{2}
f^{ij}\Big[
T_{\alpha,i}\left(v_{\beta,j} + l^\gamma w_{\beta\gamma,j}\right)
-
T_{\beta,i}\left(v_{\alpha,j} + l^\gamma w_{\alpha\gamma,j}\right)
\Big]
\,.
\end{equation}

Denoting by $w^{\alpha\beta}(q^i)$  the inverse of (\ref{w}) satisfying
\begin{equation}
w^{\,\alpha\gamma}(q^i)\,w_{\gamma\beta}(q^i) = \delta^{\,\alpha}_{\,\beta}\,,
\end{equation}
we may write $\Delta^{\alpha\beta}_{(rs)}$ for the inverse of the constraints matrix (\ref{CM}), explicitly given by
\begin{equation}\label{Delta}
\Delta^{\alpha\beta}_{(rs)} =
\displaystyle
\left[
\begin{array}{cccc}
X^{\alpha\beta} & Y^{\alpha\beta} & E^{\beta\alpha} & -w^{\alpha\beta} \\ 
-Y^{\beta\alpha} & M^{\alpha\beta} & -w^{\alpha\beta} & 0 \\
-E^{\alpha\beta} & w^{\alpha\beta} & 0 & 0 \\
w^{\alpha\beta} & 0 & 0 & 0
\end{array}
\right]
\,,
\end{equation}
with the upper index quantities
\begin{equation}\label{X}
X^{\alpha\beta} \equiv w^{\alpha\gamma}
\Big[
E_{\epsilon\gamma}w^{\epsilon\mu}\big( M_{\mu\nu}w^{\nu\xi}E_{\xi\delta} - R_{\mu\delta} \big)
+ R_{\epsilon\gamma}w^{\epsilon\mu}E_{\mu\delta} + N_{\gamma\delta}
\Big]w^{\delta\beta}
\,,
\end{equation}
\begin{equation}
Y^{\alpha\beta}
\equiv
-w^{\alpha\gamma}\left(E_{\eta\gamma}w^{\eta\epsilon}M_{\epsilon\delta} + R_{\delta\gamma}
\right)w^{\delta\beta}\,,
\end{equation}
\begin{equation}\label{Mupper}
E^{\alpha\beta} \equiv w^{\alpha\gamma} E_{\gamma\delta} w^{\delta\beta}
~~~~\mbox{and}~~~~
M^{\alpha\beta}\equiv w^{\alpha\gamma}M_{\gamma\delta}w^{\delta\beta}
\,,
\end{equation}
satisfying properties
\begin{equation}
X^{\beta\alpha} = - X^{\alpha\beta}  ~~\mbox{and}~~ M^{\beta\alpha}=-M^{\alpha\beta} \,.
\end{equation}
The previous equations and properties are essential to compute the systems's Dirac brackets algebraic structure in its most general form.
In fact, given two arbitrary functions $F=F(q^i,p_i,l^\alpha,\pi_\alpha)$ and $G=G(q^i,p_i,l^\alpha,\pi_\alpha)$, their Dirac Bracket is a third phase space function defined as
\begin{equation}\label{DB}
\{F, G\}^{\boldsymbol{*}} = \{F, G\} - \sum_{r,s=0}^3\{F, \chi_{(r)\alpha}\}\Delta^{\alpha\beta}_{(rs)}\{\chi_{(s)\beta}, G\}
\,,
\end{equation}
from which, using the constraint equations (\ref{Xi0}), (\ref{Xi1}) to (\ref{Xi3}) and the inverse matrix (\ref{Delta}), the non-null Dirac Brackets among the fundamental phase space variables can be directly obtained as
\begin{equation}\label{qpDB}
\{q^i,p_j\}^{\boldsymbol{*}} = \delta^i_j  - f^{ik} T_{\alpha,k}w^{\alpha\beta}T_{\beta,j}
\,,
\end{equation}
\begin{eqnarray}\label{ppDB}
\{p_i,p_j\}^{\boldsymbol{*}} &=& w^{\alpha\gamma}M_{\gamma\delta}w^{\delta\beta}T_{\alpha,i}T_{\beta,j}
\nonumber\\&&
+\, w^{\alpha\beta}
\left[
{\left(f^{kl}T_{\alpha, l}\right)}_{,i}T_{\beta, j}
-
{\left(f^{kl}T_{\alpha, l}\right)}_{,j}T_{\beta, i}
\right]
p_k
\,,
\end{eqnarray}
\begin{eqnarray}\label{llDB}
\{l^\alpha,l^\beta\}^{\boldsymbol{*}} &=& 
X^{\alpha\beta} \,,
\end{eqnarray}
\begin{equation}\label{qlDB}
\{q^i,l^\alpha\}^{\boldsymbol{*}} = - f^{ij}T_{\beta,j}w^{\beta\gamma}E_{\gamma\delta}w^{\delta\alpha}
+Q^{ij}_\beta p_j w^{\beta\alpha}
\end{equation}
and
\begin{eqnarray}\label{plDB}
\{p_i,l^\alpha\}^{\boldsymbol{*}} &=&
T_{\beta,i}Y^{\beta\alpha} + p_k \left( f^{kl}T_{\beta,l}\right)_{,i}E^{\beta\alpha}
\nonumber\\&&
-\,w^{\alpha\beta}\left[
\frac{1}{2}Q^{kl}_{\beta,i}p_k p_l - v_{\beta,i} - l^\gamma w_{\beta\gamma,i}
\right]
\,.
\end{eqnarray}

\section{Symplectic Analysis}
In the last section, we have succeeded in calculating the Dirac Brackets algebra in phase space, which rightfully substitute the Poisson Brackets for the current prototypical second class system, taking into account its constraint structure.  
However, it is well known that, by treating all constraints on the same ground, the Dirac Bergmann algorithm can lead to unnecessary cumbersome relations coming, for instance, from unphysical variables actually playing the role of Lagrange multipliers \cite{Neto:2022jgl}.  That issue can be circumvented by means of the symplectic Faddeev-Jackiw approach \cite{Faddeev:1988qp} supplemented by the BarcelosNeto-Wotzasek algorithm \cite{BarcelosNeto:1991kw, Wotzasek:1992wf, Wotzasek:1994ck} -- the FJBW formalism --, directly leading to the {\it Faddeev-Jackiw Brackets} among the physical variables, as we show below.  

In terms of the FJBW formalism \cite{Faddeev:1988qp, BarcelosNeto:1991kw, Wotzasek:1992wf, Wotzasek:1994ck}, we start by rewriting the Lagrangian (\ref{L}) in first-order in the time derivatives as
\begin{equation}\label{L0}
L^{(0)} = {\dot{q}}^i p_i - W^{(0)}
\,,
\end{equation}
with
\begin{equation}
W^{(0)}
\equiv
\frac{1}{2}f^{ij}p_i p_j + V(q^k) + l^\alpha T_\alpha (q^k)
\end{equation}
and identify the level zero symplectic variables as
\begin{equation}\label{xi0}
\xi_a^{(0)} = (q^i, p_i, l^\alpha)
\,.
\end{equation}
The superscript  ${}^{(0)}$ is used to denote the FJBW iterative process step or level.
Associated to the first-order Lagrangian (\ref{L0}), corresponding to the symplectic variables (\ref{xi0}), we have the initial zero-level symplectic matrix
\begin{equation}\label{f0}
f_{ab}^{(0)} = 
\left(
\begin{array}{ccc}
0&-\delta^i_j&0
\\
\delta^i_j&0&0
\\
0&0&0
\end{array}
\right)
\,,
\end{equation}
whose singularity signals we are in face of a constrained system.  The indexes $a,b$ in equation (\ref{f0}) run, accordingly, along the symplectic variables of the corresponding level -- in the current zero-level case precisely given by (\ref{xi0}).  Using the zero modes
\begin{equation}
v^{(0)}_\alpha = 
\left(
\begin{array}{ccc}0&0&{-\delta}_\alpha^\beta\end{array}
\right)\,
\end{equation}
we extract a first set of $m$ constraints given by
\begin{equation}\label{Omega0}
\Omega^{(0)}_\alpha \equiv v^{(0)}_\alpha~ \left(\begin{array}{c}\displaystyle\frac{\partial W^{(0)}}{\partial \xi_a^{(0)} }\end{array}\right)
= \left(
\begin{array}{ccc}0&0&{-\delta}_\alpha^\beta\end{array}
\right) ~
\left(\begin{array}{c}\displaystyle\frac{\partial W^{(0)}}{\partial q^i }\\\displaystyle\frac{\partial W^{(0)}}{\partial p_i }\\-T_\beta\end{array}\right)
=T_\alpha(q^k)
\,.
\end{equation}
Proceeding to the next step, level one in the FJBW algorithm, we introduce $m$ Lagrange multipliers $\lambda^\alpha$ and redefine the symplectic variables and Lagrangian function respectively as
\begin{equation}
\xi_a^{(1)} = (q^i, p_i, \lambda^\alpha)
\end{equation}
and
\begin{equation}\label{L1}
L^{(1)} = {\dot{q}}^i p_i +T_\alpha {\dot{\lambda}}^\alpha - W^{(1)}
\,,
\end{equation}
with
\begin{equation}
W^{(1)}
\equiv
\frac{1}{2}f^{ij}p_i p_j + V(q^k)
\,. 
\end{equation}
Hence, corresponding to the level one fist-order Lagrangian (\ref{L1}), we update the symplectic matrix (\ref{f0}) to
\begin{equation}\label{f1}
f_{ab}^{(1)} = 
\left(
\begin{array}{ccc}
0&-\delta^j_i&T_{\beta,i}
\\
\delta^i_j&0&0
\\
-T_{\alpha,j}&0&0
\end{array}
\right)
\,.
\end{equation}
The matrix (\ref{f1}) is still singular, signaling that, in the scope of the FJBW algorithm, there are more constraint relations to be extracted.  In fact, by considering now the $m$ zero modes
\begin{equation}
v^{(1)}_\alpha = 
\left(
\begin{array}{ccc}0&T_{\alpha,i}&{\delta}_\alpha^\beta\end{array}
\right)\,
\end{equation}
we obtain $m$ constraints more
\begin{equation}\label{Omega1}
\Omega^{(1)}_\alpha \equiv v^{(1)}_\alpha~ \left(\begin{array}{c}\displaystyle\frac{\partial W^{(1)}}{\partial \xi_a^{(1)} }\end{array}\right)
= f^{ij}T_{\alpha,i}p_j \,.
\end{equation}
Introducing new Lagrange multipliers $\eta^\alpha$ corresponding to (\ref{Omega1}), redefining the level two symplectic variables as 
\begin{equation}
\xi_a^{(2)} = (q^i, p_i, \lambda^\alpha, \eta^\alpha)
\end{equation}
and proceeding along the same previous lines,
we finally obtain a non-singular symplectic matrix given by
\begin{equation}\label{f2}
f_{ab}^{(2)} = 
\left(
\begin{array}{cccc}
0&-\delta^j_i&T_{\beta,i}&(f^{kl}T_{\beta,k})_{,i}p_l
\\
\delta^i_j&0&0&f^{ik}T_{\beta,k}
\\
-T_{\alpha,j}&0&0&0
\\
-(f^{kl}T_{\alpha,k})_{,j}p_l&-f^{jk}T_{\alpha,k}&0&0
\end{array}
\right)
\,.
\end{equation}
Note that (\ref{f2}) represents a $2(n+m)\times2(n+m)$ square matrix whose inverse can be readily computed and explicitly written as 
\begin{equation}\label{CMI}
{f^{ab}} =
\displaystyle 
{\left(
\begin{array}{cccc}
0 & \delta^i_j  - f^{ik} T_{\alpha,k}w^{\alpha\beta}T_{\beta,j} & -f^{ik}T_{\gamma,k}w^{\gamma\beta} & 0 \\ 
 -\delta^j_i  + f^{jk} T_{\alpha,k}w^{\alpha\beta}T_{\beta,i} & \Xi_{ij} & \Theta_i^{\,\,\beta}&  -T_{\gamma,i}w^{\gamma\beta}\\
f^{jk}T_{\gamma,k}w^{\gamma\alpha}  & -\Theta_{\,\,j}^{\alpha} &M^{\alpha\beta} &  - w^{\alpha\beta}\\
0 & T_{\gamma,j}w^{\gamma\alpha} & w^{\alpha\beta} & 0
\,
\end{array}
\right)}
\end{equation}
with
\begin{equation}
\Xi_{ij}\equiv  T_{\alpha,i}T_{\beta,j} M^{\alpha\beta} +
w^{\alpha\beta}
\left[
{\left(f^{kl}T_{\alpha, l}\right)}_{,i}T_{\beta, j}
-
{\left(f^{kl}T_{\alpha, l}\right)}_{,j}T_{\beta, i}
\right]
p_k
\,,
\end{equation}
\begin{equation}
\Theta_i^{\,\,\beta} \equiv  T_{\gamma,i}M^{\gamma\beta}+(f^{kl}T_{\gamma,k})_{,i}p_lw^{\gamma\beta}
\,,
\end{equation}
and $M^{\alpha\beta}$ above as in equations (\ref{Mupper}) and (\ref{M}).  The antisymmetric coordinate-dependent matrix $f^{ab}$ in equation (\ref{CMI}) represents the core of the FJBW formalism and captures the whole symplectic structure of the geometric dynamics restriction.

Indeed, from the inverse symplectic matrix (\ref{CMI}), we may directly read the Faddeev-Jackiw brackets as
\begin{equation}
\{q^i,p_j\}_{FJ} = \delta^i_j  - f^{ik} T_{\alpha,k}w^{\alpha\beta}T_{\beta,j}
\end{equation}
and
\begin{equation}
\{p_i,p_j\}_{FJ} = \Xi_{ij}
\end{equation}
which rightfully coincide with the fundamental Dirac ones obtained in the previous section.

\section{Application Examples}
In order to better clarify the previous main arguments and further exemplify our results, in this closing section, we show that the proposed PS can describe relevant trending models in physics by directly applying its concepts to two sample systems.  We shall thus see last sections' general formulae at work producing significant calculation short-cuts in their mathematical description within the proper constraints hypersurfaces.  The first application corresponds to a toroidal geometry which has been explored in many quantum field theory contexts and has been recently used in the construction of cosmological models considering the torus as a possible shaping to our universe to explain current observational data \cite{Nejad:2015rfa}, while the second one, related to Lorentz symmetry violations in quantum field theory \cite{Colladay:1998fq, Kostelecky:2003fs} concerns to an interaction between a scalar and a vector fields through a bumblebee model. 

\subsection*{Toroidal Geometry}
Particle motion on a torus as a quantum system has been largely discussed in various references from which we mention \cite{Hong:2003jp, Kumar:2014fca, Pandey:2017kwc, S:2021zob}.  Quantum fields on a torus can be seen for instance in \cite{Datta:2014zpa, Huang:2021zko, Filho:2021kso, Filho:2022tfk, Aguilera-Damia:2023jyc}.  The simplest realization sees the torus as a two-dimensional surface embedded in three dimensional space which can be described by toroidal coordinates $(\eta,\theta,\phi)$  defined through
\begin{equation}
x = (b+\eta\sin\theta)\cos\phi,
\quad y = (b+\eta\sin\theta)\sin\phi,
\quad z = \eta\cos\theta
\,.
\end{equation}
The restriction of motion to the torus surface can be naturally achieved by the Lagrangian function
\begin{eqnarray}
L = \frac{{\mu} {\dot \eta}^2}{2}+\frac{ \mu \eta^2 {\dot \theta}^2 }{2}+\frac{\mu}{2} (b+\eta \sin\theta)^2{\dot\phi}^2 - V(\eta,\theta,\phi)- l (\eta-a) \,,
\label{lagft}
\end{eqnarray}
with corresponding canonical Hamiltonian
\begin{eqnarray}
H = \frac{p^2_\eta}{2\mu}+\frac{p^2_\theta}{2\mu \eta^2}+\frac{p^2_\phi}{2\mu(b+\eta\sin\theta)^2} + V(\eta,\theta,\phi)+ l (\eta-a)\,,
\label{Ham}
\end{eqnarray}
where $\mu$ denotes the particle mass, whilst $a$ and $b$ stand for the radii of, respectively, the torus circular cross-section and its main circumference.  The potential $V$ acknowledges additional dynamics to take place within the torus surface. The extra variable $l$ plays the role of a Lagrange multiplier function.  Confronting equations (\ref{lagft}) and (\ref{Ham}) above to the corresponding ones for the general PS in expressions (\ref{L}) and (\ref{Hc}), we consider $n=3$ and $m=1$ and immediately identify
\begin{equation}
f^{ij}=\left(
\begin{array}{ccc}
f^{\eta\eta}&0&0\\
0&f^{\theta\theta}&0\\
0&0&f^{\phi\phi}
\end{array}
\right)
=
\left(
\begin{array}{ccc}
\mu^{-1}&0&0\\
0&\mu^{-1}\eta^{-2}&0\\
0&0&\mu^{-1}{(b + \eta \sin\theta)^{-2}}
\end{array}
\right)
\,,
\end{equation}
for the invertible, coordinate dependent, symmetric matrix $f^{ij}$ corresponding to the general case in equation (\ref{2}).   

Since $m=1$, we have a total of four second-class Dirac-Bergmann constraints which can be directly obtained from equations (\ref{Xi0}), (\ref{Xi1}), (\ref{Xi2}) and (\ref{Xi3}) as
\begin{equation}
\chi_0 = \pi_l\,,~~~~\chi_1=\eta-a\,,~~~~\chi_2= \mu^{-1}{p_\eta}
\end{equation}
and
\begin{equation}\label{chi3T}
\chi_3= \frac{1}{2}\left\{Q^{\theta\theta}p^2_\theta + Q^{\phi\phi}p^2_\phi\right\} - l\omega  = {\mu^{-1}}\left\{\frac{p^2_\theta}{\mu \eta^3} + \frac{p^2_\phi \sin\theta}{\mu(b + \eta\sin\theta)^3} -\frac{\partial V}{\partial\eta} - l \right\}\,,
\end{equation}
with the general coordinate-dependent quantities $w_{\alpha\beta}$ and $ Q^{ij}$, as defined in equations (\ref{w}) and (\ref{Qij}), respectively given by 
\begin{equation}
w=\mu^{-1}
\end{equation}
and
\begin{equation}
Q^{ij} = 
\displaystyle\left(
\begin{array}{ccc}
Q^{\eta\eta} & Q^{\eta\theta} & Q^{\eta\phi}\\
Q^{\theta\eta} & Q^{\theta\theta} & Q^{\theta\phi}\\
Q^{\phi\eta} & Q^{\phi\theta} & Q^{\phi\phi}
\end{array}
\right)
=
\left(
\begin{array}{ccc}
0 & 0 & 0\\
0 & {2}{\mu^{-2}\eta^{-3}} & 0\\
0 & 0 & \displaystyle\frac{2\mu^{-2}\sin\theta}{\left(b+\eta\sin\theta\right)^3}
\end{array}
\right)
\,.
\end{equation}

Furthermore, from equations (\ref{D5}) and (\ref{DR}), we have
\begin{equation}
R^{ij}=\displaystyle\left(
\begin{array}{ccc}
R^{\eta\eta} & R^{\eta\theta} & R^{\eta\phi}\\
R^{\theta\eta} & R^{\theta\theta} & R^{\theta\phi}\\
R^{\phi\eta} & R^{\phi\theta} & R^{\phi\phi}
\end{array}
\right)
= \frac{3}{\mu^3}
\left(
\begin{array}{ccc}
0 & 0 & 0\\
0 & \eta^{-4} & 0\\
0 & 0 & \displaystyle \frac{\sin^2\theta}{(b+\eta\sin\theta)^4}
\end{array}
\right)
\end{equation}
and
\begin{equation}
 R = p_\theta p_\theta R^{\theta\theta} + p_\phi p_\phi R^{\phi\phi} + f^{\eta\eta}T_{,\eta}v_{,\eta}= \frac{3}{\mu^3}\left\{\frac{p^2_\theta}{\eta^4} + \frac{p^2_\phi {\sin^2\theta}}
{(b + \eta\sin\theta)^4} \right\} + \frac{1}{\mu^2}\frac{\partial^2 V}{\partial \eta^2}
\,.
\end{equation}
Knowledge of the previous strucutre and general results (\ref{qpDB}) to (\ref{plDB}) then leads to the non-null fundamental Dirac brackets
\begin{equation}
\{\theta, p_\theta\}^* = \{\phi, p_\phi\}^* = 1\,,
\end{equation}
\begin{equation}
\{\theta, l\}^* = \frac{2p_\theta}{\mu\eta^3}\,,~~~~\{\phi, l\}^* = \frac{2\sin\theta p_\phi}{\mu(b+\eta\sin\theta)^3}\,,
\end{equation}
\begin{equation}
\{p_\theta, l\}^* = \frac{\cos\theta\left( 2\eta\sin\theta-b \right) p_\phi^2}{\mu\left(b+\eta\sin\theta\right)^4}+{\mu}\frac{\partial^2 V}{\partial\eta\partial\theta}\,,~~~~
\{p_\phi, l\}^* = \mu \frac{\partial^2 V }{\partial \eta\partial \phi}\,.
\end{equation}

Concerning the symplectic approach, the FJ constraints in equations (\ref{Omega0}) and (\ref{Omega1}) reduce to 
\begin{equation}\label{Omegas1}
\Omega^{(0)} = \eta-a  ~~~~\mbox{and}~~~~\Omega^{(1)} = \mu^{-1}p_\eta
\,,
\end{equation}
while the final inverse symplectic matrix can be read directly from (\ref{CMI}) as
\begin{equation}
{f^{ab}} = 
\left(
\begin{array}{cccccccc}
0&0&0&0&0&0&-1&0
\\
0&0&0&0&1&0&0&0
\\
0&0&0&0&0&1&0&0
\\
0&0&0&0&0&0&0&-\mu
\\
0&-1&0&0&0&0&0&0
\\
0&0&-1&0&0&0&0&0
\\
1&0&0&0&0&0&0&-\mu
\\
0&0&0&\mu&0&0&\mu&0
\end{array}
\right)
\,.
\end{equation}

Hence, the FJ brackets can be seen to exactly match the Dirac ones as well as previous results in the literature.  Therefore it is clear that, once we have the geometric characterization of the system from the matrices $f^{ij}$ and $w^{\alpha\beta}$, the canonical and Faddeev-Jackiw structures are well-defined and can be short-cut calculated from the PS general expressions.

\subsection*{Bumblebee Model}
Our second application example stems from a toy model which spontaneously breaks Lorentz symmetry.  The investigation of Lorentz violations in general relativity has produced a class of vector field theories
known as bumblebee models \cite{Kostelecky:2003fs, Nambu:1968qk, Kostelecky:1989jp, Kostelecky:1989jw, Bluhm:2004ep, Bailey:2006fd, Bluhm:2006im} which led to the development of the gravity sector of the Standard Model Extension (SME).  
In order to pass from a one dynamical parameter description $(1+0)$ to a usual field theory in $D$ space-time dimensions, we consider the de Witt notation, in which the implicit repeated index summations in the previous equations, such as in (\ref{L}) and (\ref{Hc}),  also include proper continuous space integrations.  In this way, we define our model by the action functional
\begin{equation}\label{S}
S[\phi,B^\mu]=-\frac{1}{2}\int d^Dx\,
\left[
\partial_\mu B_\nu \partial^\mu B^\nu
+\phi\left(B_\mu B^\mu - a^2 \right)
\right]
\,,
\end{equation}
describing the interaction between a scalar and a vector fields, respectively denoted $\phi$ and $B_\mu$, living in a $D$-dimensional Minkowsky space-time.\footnote{We consider a $D$-dimensional Minkowski space-time with flat metric $\eta^{\mu\nu}$ and Lorentz indexes running from $0$ to $D-1$.}  The real parameter $a$ in (\ref{S}) leads to a spontaneous Lorentz symmetry breaking as the field $B_\mu$ must favor a specific direction in space-time to produce a corresponding non-null expectation value.
In order to associate to the general PS (\ref{L}),
the Lagrangian density corresponding to action (\ref{S}) can be written as
\begin{equation}
{\cal L}=-\frac{1}{2}\eta_{\mu\nu}\dot{B}^\mu\dot{B}^\nu
-{\cal V}(B^\mu)
-\frac{1}{2}\phi\left(B_\mu B^\mu - a^2 \right)
\,,
\end{equation}
with a potential density function
\begin{equation}\label{V(B)}
{\cal V}(B^\mu)\equiv
-\frac{1}{2}\nabla B_\mu\cdot\nabla B^\mu
\,.
\end{equation}
Hence, in accordance with the general notation employed in the PS, we have
\begin{equation}
T\equiv \frac{1}{2}\left(B_\mu B^\mu - a^2 \right)
\end{equation}
and
\begin{equation}
\label{Tmu}
T_\mu(x,y)|_{y^0=x^0}=B_\mu(x)\delta^{(D-1)}(\mathbf{x}-\mathbf{y})
\,.
\end{equation}

Performing a Legendre transformation to phase space, from  equation (\ref{Hc}), we have
\begin{equation}\label{75}
H=\int\,d^{D-1}x
\left[
\frac{1}{2}{(\Pi^i)}^2-\frac{1}{2}{(\Pi^0)}^2
+{\cal V}(B^\mu)
+\frac{1}{2}\phi\left(B_\mu B^\mu - a^2 \right)
\right]
\,,
\end{equation}
with canonical momenta $\pi_\phi$ and $\Pi_\mu$ for $\phi$ and $B^\mu$, and the identification
\begin{equation}\label{feta}
f^{ij}(q^k)\longrightarrow-\eta^{\mu\nu}\delta^{(D-1)}(\mathbf{x}-\mathbf{y})
\,.
\end{equation}
The key symmetric matrix (\ref{w}) can be calculated from (\ref{Tmu}) and (\ref{feta}) as
\begin{equation}
w(x,y)|_{y^0=x^0}=-B_\mu(x) B^\mu(x) \, \delta^{(D-1)}{(\mathbf{x}-\mathbf{y})}
\,,
\end{equation}
and the general expressions  (\ref{Qij}) and (\ref{v}) are reduced to
\begin{equation}\label{Qmunu}
Q^{\mu\nu}(x,y,z)|_{z^0=y^0=x^0}=2\eta^{\mu\nu}\delta^{(D-1)}(\mathbf{x}-\mathbf{y})\delta^{(D-1)}(\mathbf{y}-\mathbf{z})
\end{equation}
and
\begin{equation}
v = - B_\mu \nabla^2 B^\mu
\,.
\end{equation}
The constraint structure given by  (\ref{Xi0}), (\ref{Xi1}), (\ref{Xi2}) and (\ref{Xi3}) can be explicitly written as 
\begin{equation}\label{Bconsts}
\begin{aligned}
{\chi}_{(0)}&=\pi_\phi
\,,~~~
\chi_{(1)}=\frac{1}{2}\left(B_\mu B^\mu - a^2 \right)
\,,~~~
{\chi}_{(2)}=-\Pi_\mu B^\mu
\,,\\
{\chi}_{(3)}&= \Pi_\mu \Pi^\mu + \phi B_\mu B^\mu + B_\mu
\nabla^2 B^\mu
\,,
\end{aligned}
\end{equation}
and can be checked to indeed constitute a second-class set.

The Dirac brackets can be readily obtained from the general expressions (\ref{qpDB}) to (\ref{plDB}) as
\begin{eqnarray}
\{B^\mu, \Pi_\nu\}^* &=& \delta^\mu_\nu - B^\mu B_\nu (B_\lambda B^\lambda)\,,\\
\{B^\mu, B_\nu\}^* &=&  0 \,,\\
\{\Pi_\mu, \Pi_\nu\}^* &=& \left(B_\lambda B^\lambda\right)^{-1} \left(\Pi_\mu B_\nu  -   \Pi_\nu B_\mu \right)\,.
\end{eqnarray}

In terms of the FJBW approach, we find the true constraints from relations (\ref{Omega0}) and (\ref{Omega1}) as
\begin{equation}
\Omega^{(0)}=\frac{1}{2}\left(B_\mu B^\mu - a^2 \right)
~~~\mbox{and}~~~
\Omega^{(1)}=-\Pi_\mu B^\mu\,,
\end{equation}
while the two symplectic matrices in equations (\ref{f2}) and (\ref{CMI}) are respectively given by
\begin{equation}
f_{ab}^{(2)} = 
\left(
\begin{array}{cccc}
0&-\delta^\nu_\mu&B_\mu&-\Pi_\mu
\\
\delta^\mu_\nu&0&0&-B^\mu
\\
-B_\nu&0&0&0
\\
\Pi_\nu&B^\nu&0&0
\end{array}
\right)
\end{equation}
and
\begin{equation}\label{f2-Ex2}
{f^{ab}} =
\displaystyle (B_\lambda B^\lambda)^{-1}
{\left(
\begin{array}{cccc}
0 & \delta_\nu^\mu B_\lambda B^\lambda - B^\mu B_\nu  &  -B^\mu & 0 \\  
-\delta_\mu^\nu B_\lambda B^\lambda + B^\nu B_\mu  &    \Pi_\mu B_\nu  -   \Pi_\nu B_\mu  &{\Pi_\mu} &  B_\mu \\
B^\nu & - {\Pi_\nu}& 0 &  1\\
0 & - B_\nu & - 1 & 0
\,
\end{array}
\right)}
\end{equation}
for a corresponding level-2 symplectic variables set
\begin{equation}
\xi_a^{(2)} \equiv (B^\mu, \Pi_\mu, \lambda, \eta)
\,.
\end{equation}
From (\ref{f2-Ex2}), the Faddeev-Jackiw bracket structure is seen to coincide with the Dirac brackets for the fundamental variables.  As we have seen, after the thorough PS constraints and symplectic structures analysis, the corresponding results for specific cases easily follow from the general one.

\section{Conclusion}
We have proposed a fairly general approach for both symplectic and canonical quantizations of second-class Dirac-Bergmann systems with holonomic constraints of the form (\ref{L}).  The symplectic formalism has proven very useful for the geometric description of dynamical motion along a differential submanifold characterized by the symplectic key matrix (\ref{w}).  We have obtained the Dirac and Faddeev-Jackiw brackets in its most general form, allowing a ready application for the two sample models discussed here corresponding to a toroidal geometry and to a Lorentz breaking toy model.  The compact index notation with Einstein summation convention used here, although in principle discrete, certainly allows for a continuous interpretation including space integrations, as attested by the field theory model application in last section.  The algebraic structure introduced and developed for the PS may be directly applied to many existing models in the literature with different geometries.   Closely related to the torus, an interesting possibility to mention is  
the geometry of a torus knot \cite{Pandey:2017qom, S:2021msx, S:2023hhk, Das:2015sdx} which has found recent applications in quantum mechanics and field theory \cite{Filho:2021kso, Filho:2022tfk, Das:2015sdx}.  The symplectic structure for the torus knot can be obtained in a natural way from the PS discussed here, as a particular case.  The non-linear sigma model and similar structures used in string theory can also be approached by the PS.  In passing, it is tempting to suggest that the PS general symplectic structure could be used to generated alternative geometrical shapes for cosmological models along the lines of references 
\cite{Nejad:2015rfa, Stevens:1993zz, Cornish:2003db, Vaudrevange:2012da}.
Further generalizations and properties at quantum level  of the PS discussed here, as well as other applications in field theory and relativistic systems of the form \cite{ThibesCamilaSuzy}, are current under analysis \cite{Thibes}.

\end{document}